\documentclass[aps,prl,10pt,twocolumn,superscriptaddress,balancelastpage,showpacs,reprint,english]{revtex4-1}
\usepackage[T1]{fontenc}
\usepackage[latin9]{inputenc}
\usepackage{amsmath}
\usepackage{amssymb}
\usepackage{graphicx}
\usepackage{color}

\makeatletter

\newcommand{\qq}{\begin{eqnarray}}
\newcommand{\qqq}{\end{eqnarray}}

\newcommand{\p}{\partial}

\newcommand{\bfx}{\mathbf{x}}

\newcommand{\bfr}{\mathbf{r}}

\newcommand{\red}{\color{red}}

\pdfpageheight\paperheight
\pdfpagewidth\paperwidth

\makeatother

\usepackage{babel}

\begin{document}

\title{ Classical Nucleation Theory for Active Fluid Phase Separation}
\author{M.E. Cates}
\affiliation{DAMTP, Centre for Mathematical Sciences, University of Cambridge, Wilberforce Road, Cambridge CB3 0WA, UK}
\author{C. Nardini}
\affiliation{Service de Physique de l'Etat Condens\'e, CEA, CNRS Universit\'e Paris-Saclay, CEA-Saclay, 91191 Gif-sur-Yvette, France}
\affiliation{Sorbonne Universit\'e, CNRS, Laboratoire de Physique Th\'eorique de la Mati\`ere Condens\'ee, 75005 Paris, France}

\date{\today}

\begin{abstract}
Classical nucleation theory (CNT), linking rare nucleation events to the free energy landscape of a growing nucleus, is central to understanding phase-change kinetics in passive fluids. Nucleation in non-equilibrium systems is much harder to describe because there is no free energy, but instead a dynamics-dependent quasi-potential that typically must be found numerically. 
Here we extend CNT to a class of active phase separating systems governed by a minimal field-theoretic model (Active Model B+). In the small noise and supersaturation limits that CNT assumes, we compute analytically the quasi-potential, and hence nucleation barrier, for liquid-vapor phase separation. Crucially to our results, detailed balance, although broken microscopically by activity, is restored along the instanton trajectory, which in CNT involves the nuclear radius as the sole reaction coordinate. \end{abstract}

\maketitle


Active fluids dissipate energy at the microscale: each constituent particle extracts energy from the environment and uses it to overcome frictional or viscous drag and create motion~\cite{ramaswamy2017active,Marchetti2013RMP}. Phase separation is ubiquitous in active systems: as in equilibrium, it can stem from attractive forces~\cite{redner2013reentrant,alarcon2017morphology}, such as adhesion which underlies compartmentalization in biological tissues~\cite{beysens2000cell,nagai2001dynamic,sussman2018soft}. Phase separation can also emerge for purely repulsive motile particles~\cite{Tailleur:08,Cates:15,Fily:12}, a situation with no equilibrium counterpart. Recently it was shown that active phase separation can displays non-equilibrium features at the macroscopic scale, such as negative surface tensions~\cite{fausti2021capillary,tjhung2018cluster,bialke2015negative}, mesoscopic currents in the steady state~\cite{tjhung2018cluster,stenhammar2014phase,caporusso2020micro,singh2019hydrodynamically}, or highly dynamical clustering~\cite{Palacci:12,Speck:13,van2019interrupted}. Below we address the simplest case where the active system undergoes bulk fluid-fluid phase separation. Although at first sight this resembles closely the equilibrium case~\cite{Speck2014PRL,Tailleur:08,fodor2016far,Brader:15,Maggi:15,szamel2016theory,nardini2017entropy}, detailed balance remains broken mesoscopically in the presence of density gradients~\cite{nardini2017entropy,martin2021statistical}. In phase-field type models, the resulting interfacial activity alters the binodal densities at coexistence~\cite{Wittkowski14,solon2018generalized}. It must likewise be accounted for to properly define the pressure in particle-based models~\cite{solon2015pressure}.

A crucial feature of phase-separating systems is homogeneous nucleation, a rare event causing the formation of a distinct phase by growth of a nucleus within the bulk of a metastable parent phase. This growth is driven by noise until a critical radius is reached whereafter it proceeds spontaneously.
In passive fluids, Classical Nucleation Theory (CNT)~\cite{debenedetti2021metastable,oxtoby1992homogeneous} states that the probability of nucleating a liquid droplet in a vapor with supersaturation $\epsilon$ is given, within the large deviations limit of low temperature $T$, by  $\mathbb{P} \asymp \exp\left(-U_{eq}(R_c)/k_BT\right)$. Here, $\asymp $ stands for logarithmic equivalence~\cite{touchette2009large} and $k_B$ is the Boltzmann constant. In three spatial dimensions, the free energy barrier is given by
\qq\label{eq:DeltaF-eq}
U_{eq}(R_c)
= \frac{4\pi}{3}\sigma_{eq} R_{c,eq}^2 + \mathcal{O}(R_c,T)\qquad d=3\,
\qqq
in terms of the critical radius is $R_{c,eq}= 2\sigma_{eq}/(f'(\phi_s)\Delta\phi-\Delta f)$ and $\sigma_{eq}$ is the surface tension of the interface. Here, $\Delta\phi=\phi_2-\phi_1$ and $\Delta f=f(\phi_2)-f(\phi_1)$ where $\phi$ is the order parameter ({\em e.g.}, particle density); $f(\phi)$ is the corresponding free-energy density; $\phi_{1,2}$ represent respectively the vapor and liquid binodals, and $\phi_s=\phi_1+\epsilon$. 
CNT holds for small supersaturation ($\epsilon\ll |\phi_{1}|$) such that the critical nucleation radius $R_c$ is large compared to the interfacial width. It assumes that the nucleus remains almost spherical, which is true for fluid-fluid phase separation in the regime just delineated. CNT equally describes nucleation of vapor from liquid by interchanging $1\leftrightarrow 2$.
The vast literature on CNT has {\em inter alia} aimed at testing it experimentally and numerically~\cite{oxtoby1992homogeneous,karthika2016review}; at improving its predictions beyond the limit of small supersaturation~\cite{cahn1959free}; at describing systems where multiple pathways to nucleation are present~\cite{wolde1997enhancement}, and at assessing the relative importance of homogeneous and heterogeneous nucleation~\cite{sear2007nucleation}.

It has been suggested that CNT might be extended to address nucleation in phase-separating active systems~\cite{richard2016nucleation,redner2016classical,levis2017active}, but there has been limited progress along these lines so far. We are aware of one study, restricted to hard-core non-Brownian particles, which assumes a nucleation pathway via single-monomer attachments, and requires fitting parameters to get quantitative agreement with simulations~\cite{redner2016classical}. Below we address instead CNT via statistical field theory. Here we will find that the standard analysis for passive systems can be extended with surprising completeness to the active case.

Classical nucleation theory is one prominent instance of large deviation theory (LDT)~\cite{touchette2009large,bertini2015macroscopic}, which addresses rare events in settings ranging from solid state physics~\cite{Bray} and physical chemistry~\cite{allen2009forward} to finance~\cite{bouchaud1998langevin}, turbulence~\cite{falkovich1996instantons,ravelet2004multistability}, and geophysical flows~\cite{schmeits2001bimodal,ragone2018computation}. 
In thermal equilibrium systems, event rates can be found from the free energy barrier, {\em e.g.} via \eqref{eq:DeltaF-eq} above
(although dynamical methods can also be used~\cite{Lutsko2012}). By working with the free energy, one also accesses the typical dynamics of the rare event: time-reversal symmetry ensures that the most probable route up the barrier (the so-called instanton path) is the time-reversal of the noiseless (relaxational) downward path~\cite{graham1987macroscopic,freidlin1998random}. 
 
The situation is very different in non-equilibrium systems such as active matter. Within LDT~\cite{touchette2009large,bertini2015macroscopic} the free energy is replaced by the quasi-potential~\cite{graham1987macroscopic,freidlin1998random}, but this is unknown {\em a prori}, and the instanton is not in general the time-reversal of the relaxational path. Computing the quasi-potential and/or instanton represents an intrinsically dynamical problem which, even in the small noise limit of LDT, is rarely achievable analytically. (Only for a few minimal models was the quasi-potential found either exactly~\cite{derrida1998exact,bodineau2005current,mallick2022exact}, or by perturbation theory~\cite{maier1993escape,tel1989nonequilibrium,bouchet2016perturbative}.) Even from a numerical perspective, studying rare events without detailed balance is much more complex than at equilibrium; dedicated algorithms developed for this task~\cite{bucklew2004introduction} include cloning~\cite{giardina2006direct,lecomte2007numerical}, instanton-based codes~\cite{heymann2008pathways,vanden2012rare,grafke2015instanton}, and other approaches~\cite{grafke2019numerical,nemoto2014computation,ferre2018adaptive,yan2022learning,zakine2022minimum}. Accordingly, while intense research into rare events in active systems was recently initiated~\cite{zakine2022minimum,fodor2022irreversibility,cagnetta2017large,yan2022learning,woillez2019activated,agranov2022,das2022direct}, this has been mainly numerical. 

In this Letter we extend CNT to active fluid phase separation, using statistical field theory. We can thereby access analytically nucleation rates and quasi-potentials for a generic class of non-equilibrium, many-body systems. This is possible because, although activity breaks detailed balance, this is restored along the instanton trajectory, which in CNT involves a single reaction coordinate (the droplet radius) with noise that we infer from the infinite-dimensional Langevin equation for the order parameter field. Our results are given for Active Model B+ (AMB+)~\cite{nardini2017entropy,tjhung2018cluster}, a canonical field theory for active phase separation. However, the analysis route just outlined should be open whenever CNT's precept of a single reaction coordinate is applicable.

In their simplest form~\cite{Wittkowski14,tjhung2018cluster,thomsen2021periodic}, statistical field theories of active phase separation only retain the evolution of a composition or density field, $\phi$. (Hydrodynamic~\cite{tiribocchi2015active,singh2019hydrodynamically} or polar~\cite{tjhung2012spontaneous} fields can be added if required.) Their construction proceeds via conservation laws, symmetry arguments, and an expansion in $\phi$ and its gradients, along lines long established for Model B, which describes passive phase separation~\cite{hohenberg1977theory,chaikin2000principles,bray2001interface}. In the active case, locally broken time-reversal symmetry implies that new non-linear terms are allowed. The ensuing minimal theory, AMB+, includes all terms that break detailed balance up to order $\mathcal{O}(\nabla^4,\phi^2)$~\cite{nardini2017entropy,tjhung2018cluster}: 
\qq
\p_t\phi&=&-\nabla\cdot\left(\mathbf{J}+\sqrt{2D}\mathbf{\Lambda}\right)\,\label{eq:AMB}\\
 {\bf J}/M &=&-\nabla \mu_\lambda  + \zeta (\nabla^2\phi)\nabla\phi\,\label{eq:AMB+J}\\
\mu_\lambda[\phi] &=& \frac{\delta \mathcal{F}}{\delta\phi} +\lambda|\nabla\phi|^2\;.\label{eq:AMB+mu}
\qqq
Here $\mathcal{F} = \int d\bfr \,\left[f(\phi) +\frac{K(\phi)}{2}|\nabla\phi|^2\right]$, with $f(\phi)$ a double-well local free energy density,
and $\mathbf{\Lambda}$ is a vector of zero-mean, unit-variance, Gaussian white noises. Below we choose unit mobility ($M=1$); set $K$ constant (though our results can be extended to any $K(\phi)>0$); assume constant noise $D$; and choose $f(\phi)$ as a quartic polynomial. These are standard simplifications for passive Model B, which is recovered, setting $D=Mk_BT$, at vanishing activity ($\lambda=\zeta=0$)~\cite{hohenberg1977theory}, and leads to \eqref{eq:DeltaF-eq}. 
As shown in~\cite{tjhung2018cluster,solon2018generalized} the explicit coarse-graining of quorum-sensing particle models leads $\zeta=0$, while non-vanishing $\zeta$ and $\lambda$ are obtained when two-body forces are included~\cite{tjhung2018cluster,vrugt2022derive}.
Note that the $\zeta$ term in \eqref{eq:AMB+J} can be written, via Helmholtz decomposition, as $-\nabla\mu_\zeta +\nabla\wedge {\bf A}$, whose second, divergenceless part does not affect the $\phi$ dynamics in \eqref{eq:AMB}. Thus we define a total chemical potential $\mu=\mu_\lambda+\mu_\zeta$, with $\mu_\zeta$ nonlocal in $\phi$~\cite{supp}. 

Let us denote by $\phi_{1,2}$ the binodal densities at which bulk vapor and liquid phases coexist. Within LDT,  these can be calculated at mean field ($D\to 0$) level; without activity this amounts to global minimization of ${\mathcal F}$.
For AMB+ they are instead found by changing variables from $\phi$ and $f$ to $\psi$ and $g$: these solve $K\p^2 \psi/\p\phi^2 = (\zeta-2\lambda)\p \psi/\p\phi$ and $\p g/\p \psi = \p f/\p \phi$, where in uniform bulk phases $\p f/\p \phi = \mu$ as defined previously.
It follows that $\psi =K \left(\exp[(\zeta-2\lambda) \phi/K] -1\right)/(\zeta-2\lambda)$~\cite{solon2018generalized,tjhung2018cluster}. In transformed variables, the binodal densities $\phi_{1,2}$ obey the usual equilibrium conditions: $\mu_1=\mu_2$ and $(\mu\psi-g)_1 = (\mu\psi-g)_2$~\cite{tjhung2018cluster,solon2018generalized}. This change of variables 
vastly simplifies the mathematical construction of phase equilibria but we show in~\cite{supp} how our main results can be found without them. 

\begin{figure}[h]\center
  \centering
    \includegraphics[width=0.7\linewidth]{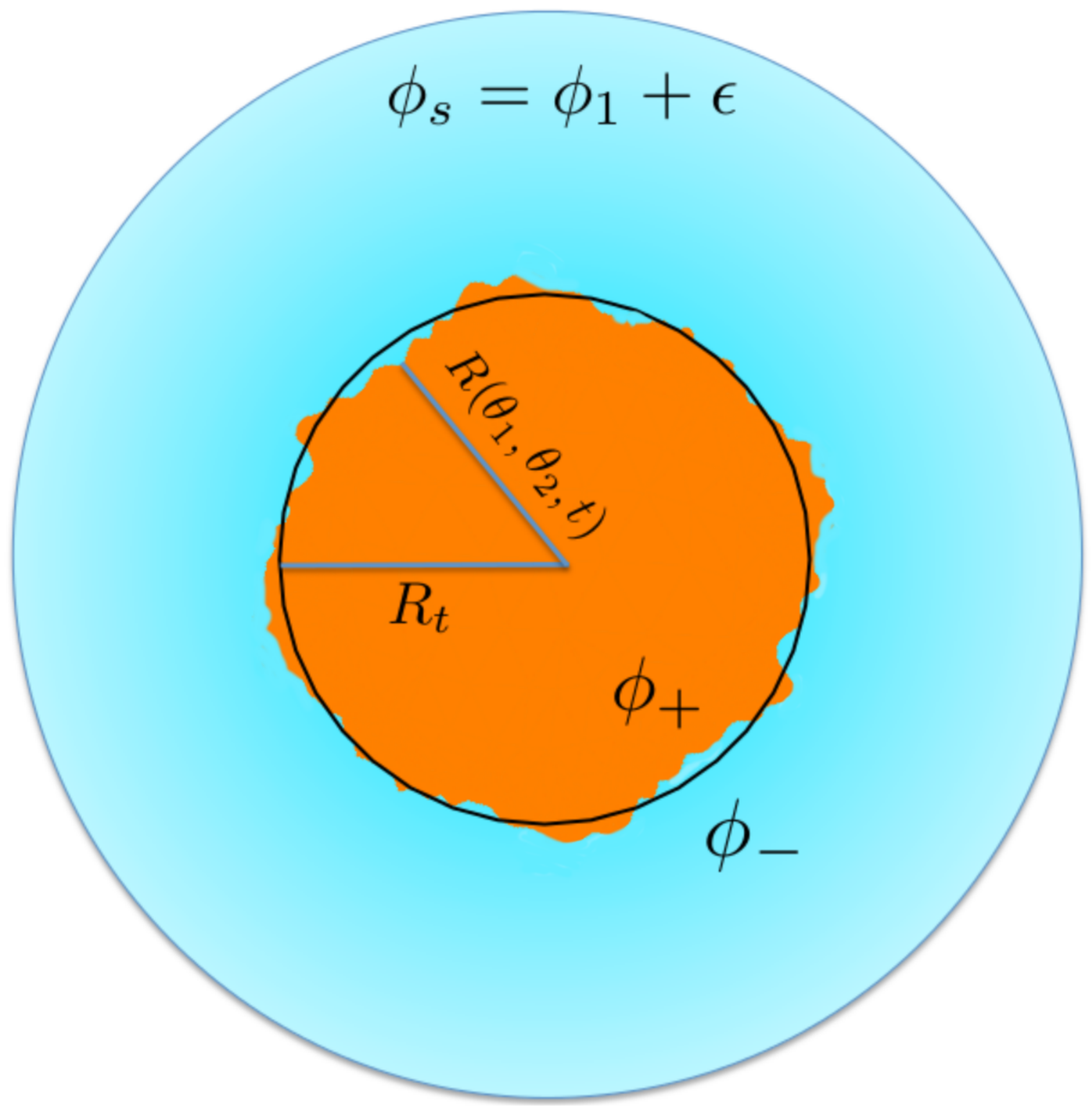}
    \caption{ A nucleating liquid (orange) droplet in a vapor (blue) environment, showing the notation used in the text. }
    \label{fig:e}
\end{figure}

We now consider, as in Figure \ref{fig:e}, the nucleation of a liquid droplet of mean radius $R_t$ in a supersaturated vapor with density at infinity $\phi_s=\phi_1+\epsilon$ (vapor-in-liquid nucleation can of course be addressed likewise). We detail our analysis in $d=3$ but our main results are valid in dimensions $d\geq 2$; the case $d=2$ involves boundary terms and we treat it separately below.
Following CNT, we assume small supersaturation, $\epsilon$; the chemical potential at infinity remains constant and equal to $f'(\phi_s)$.
The central idea is that, as in passive fluids~\cite{Onuki,Lutsko2012}, the most probable nucleation path is explored by quasi-static diffusion in the low noise limit. The locally coexisting densities outside and inside a near-spherical droplet (at distance from the interface large compared to its width $\xi$ but small compared to the droplet size) are denoted by $\phi_{\pm}$. We parameterise the droplet by its radius $R(\theta_1,\theta_2,t)=R_t+\delta R(\theta_1,\theta_2,t)$ as a function of the polar and azimuthal angles, where $\delta R$ encodes shape fluctuations at fixed volume such that $\int d\theta_1 d\theta_2 \sin\theta_1 \delta R =0$. 

In principle, three separate sources of fluctuations could contribute to the nucleation dynamics: $(i)$ fluctuations in mean droplet radius $R_t$; $(ii)$ non-spherical fluctuations of its shape encoded in $\delta R$; $(iii)$ fluctuations in the density profile $\varphi_{R_t}(r-{R})$ normal to the interface. 
However, under CNT's assumption of large $R_t$ and small noise amplitude, only fluctuations in the droplet radius are relevant for nucleation, just as in the passive case. 
To confirm this for active systems, we note first that shape fluctuations are driven only by noise -- nothing else in \eqref{eq:AMB} breaks rotational symmetry. As shown in~\cite{fausti2021capillary}, these fluctuations are resisted by a capillary-wave interfacial tension 
$\sigma_{\rm cw}= 
\int dy \, \varphi'^2(y)\left\{ K
-\zeta 
\left[\psi(y) -(\psi_1+\psi_2)/2\right]\right\}$. 
Hence $\delta R\sim \mathcal{O}(\sqrt{D/\sigma_{\rm cw}})$ and, as confirmed below, this is negligible for nucleation dynamics at small $D$. Second, fluctuations in the normal profile relax on a diffusive time-scale set by the interfacial thickness and hence decay fast compared to the diffusive growth or shrinkage of the droplet itself.
This can be formalized in the Ansatz $\phi(\bfx, t) = \varphi_{R_t}(r- R(\theta_1,\theta_2,t))$,  
\qq\label{eq:Ansatz}
\varphi_{R_t}(r- R)=
\varphi(r-R)+\frac{\varphi_1(r-R)}{R_t}+\mathcal{O}(R_t^{-2})\,
\qqq
where $\varphi$ is the stationary density profile of a flat interface and $\varphi_1$ a correction due to curvature.

We can now derive the stochastic evolution equation for the mean droplet radius $R_t$. 
We start by inserting (\ref{eq:Ansatz}) in (\ref{eq:AMB}). Working within the Stratonovich convention, integrating over the angular coordinates, and retaining only the leading orders in $R_t^{-1}$ and $\delta R$, the left hand side gives $-\varphi_{R_t}'(r-R_t) \dot{R}_t+\mathcal{O}( R_t^{-2},\delta R^2)$, where $\varphi_{R_t}'(r)\equiv \p_r \varphi_{R_t}(r)$. Denoting by $ \mu_{R_t}(r)=\mu[\varphi_{R_t}(r-R_t)]$ the chemical potential evaluated at $r$ for the profile $\varphi_{R_t}(r-R_t)$, we thus obtain
\qq\label{eq:pt-spherical}
-\varphi_{R_t}'(r-R_t) \dot{R}_t =  \nabla^2 \mu_{R_t} +\chi +\mathcal{O}( R_t^{-2},\delta R^2)\,
\qqq
where the Gaussian white noise  $\chi$ has zero mean and variance $\langle \chi(r_1,t_1)\chi(r_2,t_2)\rangle  = -2D\nabla^2(\delta(r_1-r_2)/r_1^{d-1})\delta(t_1-t_2)/S_d$, with $S_d=4\pi$ in $d=3$. In deriving the noise term we have transformed the Dirac delta from cartesian to spherical coordinates using $\delta(\bfx-\bfx')d\bfx=\delta(r-r')\delta(\theta_1-\theta_1')'\delta(\theta_2-\theta_2')/(r^2\sin\theta_1)d\bfx$. 

We next invert the Laplacian in (\ref{eq:pt-spherical}). Imposing as boundary conditions that $\mu=f'(\phi_s)$ at $r\to\infty$, and that the solution is nonsingular at $r\to 0$, we get
\qq\label{eq:derivation-int}
 \nabla^{-2}
\left[
\varphi_{R_t}' \dot{R}_t
+\chi \right]\,  
=
f'(\phi_s)-\mu_{R_t}+\mathcal{O}(R_t^{-2},\delta R^2)\,.
\qqq
Here, 
$\nabla^{-2}s$ denotes the solution $\ell(r)$ to the Poisson equation $\nabla^2\ell(r) = s(r)$, defined for all $r\in[0,\infty)$ and vanishing at infinity, which explicitly reads $\ell(r)=-\int_r^\infty dr_2 \int_0^{r_2} dr_1 \,(r_1/r_2)^{d-1}\, s(r_1)$. 

The effective evolution for the radius of the droplet $R_t$ can be found multiplying (\ref{eq:derivation-int}) by $\psi_{R_t}'$, where $\psi_{R_t}$ is the transformed variable associated with $\varphi_{R_t}$, and integrating across the interface. The right side of (\ref{eq:derivation-int}) gives
\qq\label{eq:standard-OR-int}
{(d-1)}\sigma\left\{\frac{1}{R_t}-\frac{1}{R_c}\right\} +\mathcal{O}\left({ R_t^{-2}, \delta R^2}\right)\,
\qqq
where the critical radius $R_c$ is given by 
\qq\label{eq:Rc}
R_c=\frac{ (d-1)\sigma}{f'(\phi_s)\delta\psi-\delta g}\,,
\qqq
with $\delta \psi = \psi(\phi_+)-\psi(\phi_s)$, $\delta g = g(\phi_+)-g(\phi_s)$. Here, $\sigma 
= 
\int dr \, \varphi'(r)\left\{ K\psi'(r)
-\zeta 
\left[\psi(r) -\psi({ \phi_2})\right]\varphi'(r)\right\}$ is a surface tension. Unlike in passive systems, this differs from $\sigma_{\rm cw}$; it is the tension previously encountered in our studies of Ostwald ripening of liquid droplets~\cite{tjhung2018cluster}, and can take either sign. Here we restrict to $\sigma>0$, but comment later on the case of negative $\sigma$.

Under the same procedure of multiplying by  $\psi_{R_t}'$ and integrating, then using the solution to the Poisson equation and expanding in powers of $R_t^{-1}$, the deterministic term on the left side of (\ref{eq:derivation-int}) gives~\cite{supp}
\qq\label{eq:ptRt_int}
\dot{R}_t\int dr \, \psi_{R_t}'\nabla^{-2}\varphi_{R_t}' 
=
-R_t \dot{R}_t \left[\delta\phi\delta\psi+\mathcal{O}(R_t^{-1})\right]
\qqq
where $\delta\phi=\phi_+-\phi_s$. Likewise the noise term in (\ref{eq:derivation-int}) gives a Gaussian noise whose correlations we compute in~\cite{supp}. 

Putting together these results, we obtain the stochastic dynamics for the radius of the droplet:
\qq\label{eq:R_t-final}
\dot{R}_t 
=
- \mathcal{M}(R_t) \frac{\partial U}{\partial R_t} +\sqrt{2D{\mathcal{M}}}\,\Lambda+\mathcal{O}(R_t^{-3},\delta R^2).
\qqq
Here,  $\Lambda$ is a zero-mean Gaussian noise with correlations $\langle \Lambda(t_1)\Lambda(t_2)\rangle 
= \delta(t_1-t_2)$; the effective mobility is
\qq
\mathcal{M}(R_t) &=& \frac{1}{S_d(\delta\phi)^2 R_t^d}+\mathcal{O}\left(\frac{1}{R_t^{d+1}}\right)\,;\label{eq:MR}
\qqq
and the effective potential is
\begin{equation}\label{eq:effective-R-equation-landscape}
U(R_t) = \sigma  \frac{(\phi_+-\phi_s) S_d}{\psi(\phi_+)-\psi(\phi_s)} R_t^{d-1}
\left[
1-\frac{d-1}{d}\frac{R_t}{R_c}
\right]\,.
\end{equation}
Equation \eqref{eq:R_t-final}, including crucially its noise, is the same as for a thermal Langevin particle of mobility $\mathcal{M}(R_t)$ and temperature $D$ in a potential $U(R_t)$. Detailed balance is thus restored for the dynamics of $R_t$ even though it is absent in that of $\phi({\bf x},t)$.
Indeed \eqref{eq:R_t-final} is the same equation as for the reaction coordinate $R_t$ in passive CNT for Model B, up to a change in free energy landscape $U$.

 While we have derived the above results in $d=3$, they hold for any $d> 2$. For $d=2$ boundary terms arise via the inverse Laplacian, and while (\ref{eq:Rc}, \ref{eq:R_t-final}) and (\ref{eq:effective-R-equation-landscape}) still hold, (\ref{eq:MR}) is changed to
\begin{equation}
\mathcal{M}(R_t) = \frac{1}{S_d(\delta\phi)^2 R_t^2\log({R^+}/{R_t})}+\mathcal{O}\left(\frac{1}{R_t^{3}}\right)\label{eq:MR-d=2}
\end{equation}
where $R^+$ is the upper limit of integration for $r$. For a single droplet nucleating in a circular domain of radius $L$, with $\phi=\phi_s$ at the boundary, one chooses $R^+=L$. On the other hand, for a nucleation event taking place among other droplets already undergoing coarsening, $R^+$ must be self-consistently determined from the distribution $n(R,t)$ of droplets of radius $R$ at time $t$. Just as in the passive case, this yields $(R^+)^{-1} = 2\pi \int dR \,R \,n(R,t) \,K_1(R/R^+)/K_0(R/R^+)$, where $K_{0,1}$ are modified Bessel functions of the first kind~\cite{marqusee1984dynamics,zheng1989theory}. 

Eqations (\ref{eq:R_t-final}--\ref{eq:MR-d=2}) are the key results of this Letter.  Just as for passive CNT, they are valid in the limit of large $R_t$, small noise and small supersaturation, in which $R_c$ is also large. Using the fact that $\phi_\pm= \phi_{1,2}+\mathcal{O}(R_t^{-1})$~\cite{tjhung2018cluster}, one can then replace $\delta\phi$, $\delta\psi$ and $\delta g$ with $\Delta\phi$, $\Delta\psi$ and $\Delta g$~\cite{supp}, and the critical radius reduces to $R_c = (d-1)\sigma/(f'(\phi_s)\Delta\psi-\Delta g)+\mathcal{O}(\epsilon^0, 1/(\epsilon R_t))$ or, equivalently, to $R_c=(d-1)\sigma/(\epsilon \Delta\psi f''(\phi_1)) +\mathcal{O}(\epsilon^0, 1/(\epsilon R_t))$. Moreover, although (\ref{eq:R_t-final}) contains multiplicative noise, it is equivalent in the Ito and Stratonovich interpretations, as the conversion factor enters only at order $\mathcal{O}(R_t^{-d-1})$. 

As usual in CNT, a droplet successfully nucleates upon reaching the critical radius $R_c$.  The probability (or rate) for this to happen is given at large deviation level by $\mathbb{P}(R_c)\asymp\exp(-U(R_c)/D)$, with
\qq\label{eq:nucleation-barrier-pap}
U(R_c)
=
\frac{\sigma S_d}{d {\red}} \frac{\Delta\phi}{\Delta\psi} R_c^{d-1} +\mathcal{O}(R_c^{d-2}, D)
\qqq
the quasi-potential for the critical droplet. Here we have used that $\delta R\sim \mathcal{O}(\sqrt{D/\sigma_{\rm cw}})$. Just as it should, (\ref{eq:nucleation-barrier-pap}) reduces to (\ref{eq:DeltaF-eq}) for passive Model B: without activity, $\psi\to\phi$, $\sigma\to\sigma_{eq}$ and $R_c\to R_{c,eq}$. Furthermore,  it should be noted that the nucleation barrier in eq. (\ref{eq:nucleation-barrier-pap}) takes exactly the same form as in equilibrium upon the exchange of $\sigma_{eq},R_{c,eq}$ with $\sigma \Delta\phi/\Delta\psi$ and $R_c$. 
The growth of a droplet from $R_c$ to its final radius is relaxational: therefore, the probability of observing a droplet of radius $R$ is given by $\mathbb{P}(R)\asymp\exp(-\bar{U}(R)/D)$, where $\bar{U}(R)=U(R)$ if $R<R_c$ and $\bar{U}(R)=U(R_c)$ if $R\geq R_c$. 

\begin{figure}[h]\center
  \centering
    \includegraphics[width=1.\linewidth]{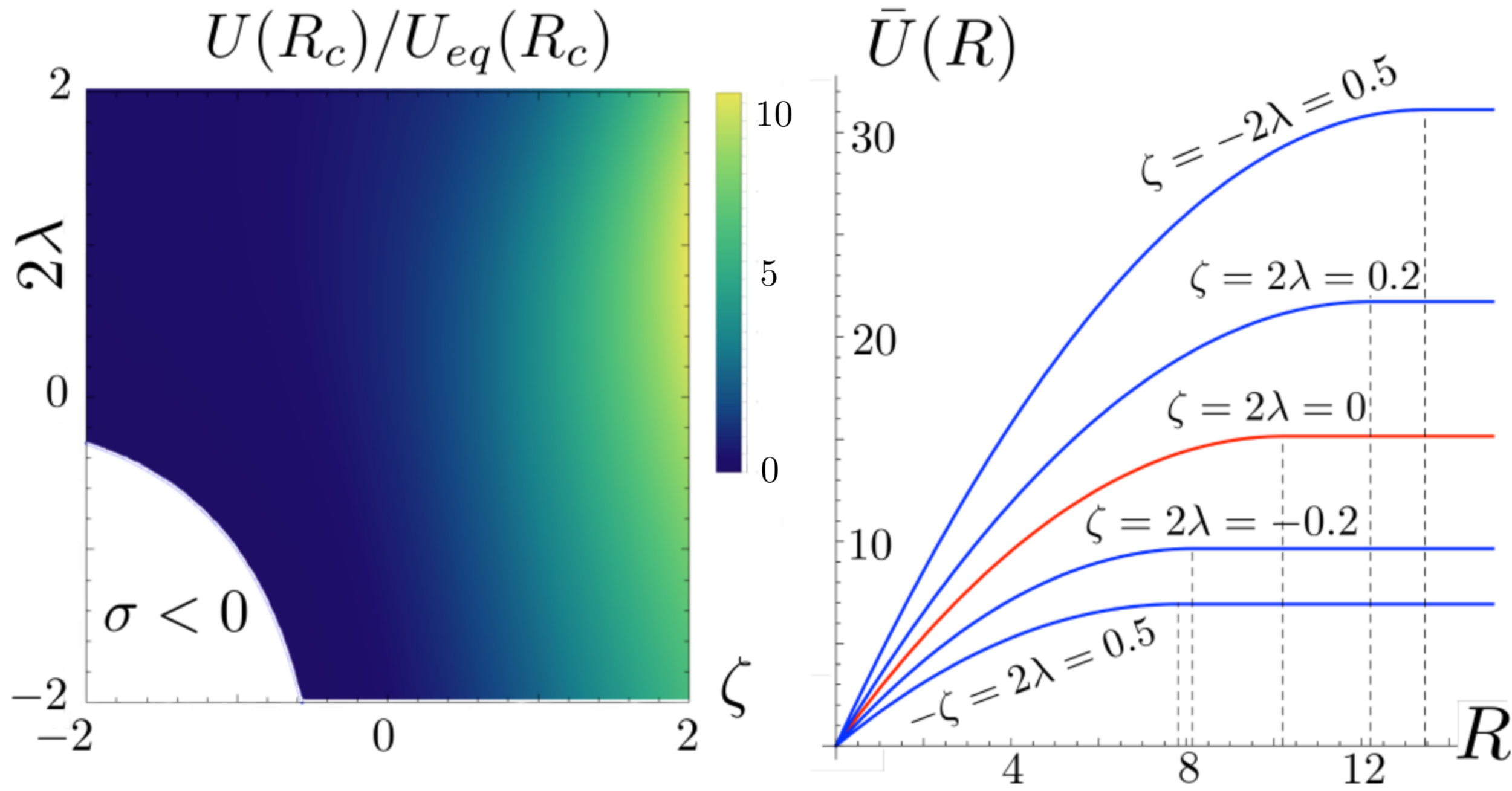}
    \caption{ (Left) Quasi-potential barrier $U(R_c)$ normalised by the equilibrium value $U_{eq}(R_c)$ to nucleate a critical liquid droplet in the two-dimensional AMB+ as a function of $\lambda$ and $\zeta$. $U(R_c)$  increases monotonically along the $\zeta$ direction but not along $\lambda$.     (Right) Quasi-potential for a droplet of radius $R$ for different values of activity. Vertical dashed lines mark the critical radius. In both panels, the supersaturation is fixed at $\epsilon=5\times 10^{-2}$ and we used $f(\phi)=(-\phi^2/2+\phi^4/4)/4$ with $K=1$.}
    \label{fig:UR}
\end{figure}

In order to explicitly compute $\bar{U}(R)$ and/ or integrate the instanton dynamics (\ref{eq:R_t-final}) we need to evaluate the surface tension $\sigma$ and the binodals. As shown in~\cite{tjhung2018cluster,supp}, these quantities can be obtained via a simple numerical procedure, with $\sigma$ found by a single numerical integral. Moreover, when $\zeta=2\lambda$, and the local free energy is of standard form $f(\phi)=-a\phi^2/2+b\phi^4/4$, the quasi-potential $\bar{U}(R)$ can be found analytically. From (\ref{eq:nucleation-barrier-pap}) we indeed have that $\psi=\phi$, $\Delta \phi=2\sqrt{a/b}$, and $\sigma=\sigma_{\rm eq}(1+\zeta/K)$, where $\sigma_{eq} = \sqrt{8Ka^3/(9b^2)}$. 
Note that any cubic term $c\phi^3/3$ in $f(\phi)$ can be absorbed by the shift $a\to a +c^2/(3b)$, $\phi\to \phi+c/3b$.
In Fig.~\ref{fig:UR} we plot $U(R_c)$ and the quasi-potential $\bar{U}(R)$ at fixed supersaturation for various $\zeta$ and $\lambda$. Unsurprisingly, the nucleation barrier is very strongly changed by activity: not only can it collapse to zero as $\sigma$ approaches negative values, but it can also be much enhanced (for positive $\zeta$). Since the barrier enters nucleation rates exponentially, our ability to compute it within CNT is a crucial step in quantitatively understanding active phase separation kinetics.

The theory derived in this Letter describes nucleation in AMB+ so long as the active surface tension $\sigma$ and the capillary wave tension $\sigma_{cw}$ are positive. Sufficiently high activity can cause capillary waves to become unstable: $\sigma_{cw}<0$ is indeed found at $\lambda,\zeta$ sufficiently large and positive~\cite{fausti2021capillary}, not shown in Fig. \ref{fig:UR}. Also, sufficiently strong activity can turn the Ostwald process into reverse while $\sigma_{cw}>0$: $\sigma<0$ is indeed found at $\zeta,\lambda$ sufficiently negative for liquid droplets (or positive for vapor bubbles)~\cite{tjhung2018cluster}. When this happens, $R_c(\epsilon)$ is a stable fixed point for the relaxational dynamics: a droplet grows (shrinks) if it is initially smaller (larger) than $R_c$. In this case the critical radius for nucleation is not $R_c$ but expected to be much smaller. Such nucleation might perhaps be captured by including the terms of order $R_t^{-3}$ in \eqref{eq:R_t-final}: we leave this idea to future work.

In conclusion, we have extended classical nucleation theory to address phase separation in active fluids. It is very unusual, if not unique, for the probability of rare fluctuations to be obtained analytically in a strongly nonequilibrium system, with continuously many degrees of freedom, as done here. This was achieved because detailed balance, although violated in the full dynamics, is restored at large deviation level for the reaction coordinate, as shown by explicit construction of the Langevin equation for droplet size \eqref{eq:R_t-final}. A key element to this was to compute the noise in this equation, which we found to be unaffected by the active terms although these drastically alter the (quasi-)potential landscape.

Our results were derived within AMB+, a canonical field theory for phase separation in active fluids. More generally, the techniques developed here could help elucidate the nucleation dynamics in more specific active models, by addressing field theories obtained by explicit coarse-graining~\cite{tjhung2018cluster,solon2018generalized,bickmann2020collective}. Our techniques might also help in understanding nucleation in the presence of non-equilibrium chemical reactions~\cite{ziethen2022nucleation,Jacobs2022}, relevant for describing the inner structure of cells~\cite{hyman2014liquid,li2020non} or when the density evolution is coupled to other slow fields~\cite{tiribocchi2015active,singh2019hydrodynamically,tjhung2012spontaneous}. More speculatively, our results might be a step towards understanding nucleation, considered a key ingredient of cancer mestastasis~\cite{basan2009homeostatic}, in biological tissues. 

Confirming numerically our analytical results poses a significant challenge; this is because CNT is valid in the regime where the nucleation barrier is much larger than the noise amplitude ($U(R_c)\gg D$) and the supersaturation is small, so that the nucleation rate is also very small and the critical radius is large. Even for passive fluids, where CNT has been widely verified both experimentally and in particle models~\cite{oxtoby1992homogeneous,karthika2016review},
we are not aware of any computational work on Model B that addressed nucleation in the CNT regime. Such a challenge can likely be addressed by employing recently developed algorithms dedicated to sample rare events in systems far from equilibrium~\cite{grafke2019numerical,nemoto2014computation,ferre2018adaptive,yan2022learning,zakine2022minimum,kikuchi2020ritz}; in fact, our exact results create a potential benchmark for these codes. A key question is whether algorithms can be created to automatically identify a low dimensional subspace of one or more reaction coordinates, without relying on detailed mechanistic analysis of the type presented above.

\begin{acknowledgments}
Work funded in part by the European Research Council under the Horizon 2020 Programme, ERC grant agreement number 740269 and by the National Science Foundation under Grant No. NSF PHY-1748958, NIH Grant No. R25GM067110 and the Gordon and Betty Moore Foundation Grant No. 2919.02. MEC was funded by the Royal Society. 
\end{acknowledgments}

\bibliographystyle{apsrev4-1}
\bibliography{biblio.bib}
\newpage
\appendix
\section{Derivation of the effective equation for the radius of the droplet}
We provide here some details on the derivation of the effective equation for the radius of the droplet. We carry on these calculations in $d=3$ but the analysis is very similar for $d\geq 2$ (for $d=2$ one must take into account the boundary terms that arise).

\subsection{Total chemical potential}
As discussed in~\cite{tjhung2018cluster} and in the main text, and AMB+ can be rewritten as $\dot{\phi} = \nabla^2 \mu -\nabla\cdot \sqrt{2D}\mathbf{\Lambda}$, where the total chemical potential $\mu=\mu_\lambda+\mu_\zeta$ is obtained via Helmholtz decomposition. This gives
\qq
\mu_{\zeta} = -
\zeta\int d\bfx_1
\nabla\cdot(\nabla^2\phi\nabla\phi)(\bfx_1) G(\bfx-\bfx_1)
\qqq
where $G$ is the Green function of the Laplacian ($G(\bfx) = -1/(4\pi |\bfx|)$ in $d=3$).
Evaluating the chemical potential $\mu$ on the spherically symmetric profile $\varphi_{R_t}(r- R_t)$, and integrating over the angular coordinates, we obtain
\qq
\mu_{R_t} &=& f'(\varphi_{R_t}) - K \varphi_{R_t}'' - \frac{(d-1)K}{r} \varphi_{R_t}' + \Big(\lambda-\frac{\zeta}{2}\Big)\varphi_{R_t}'^2  \nonumber \\
       &+& (d-1)\zeta \int_r^\infty \frac{\varphi_{R_t}'^2(r_1)}{r_1} \,dr_1 \,
\qqq
that enters in eq. (6) and (7). The derivation of eq. (8) from eq. (7) follows similar lines as those described in~\cite{tjhung2018cluster}.
\subsection{Deterministic part}
We show how to obtain Equation (10) from Equation (7). Using the solution of the Poisson equation given in the main text, we have
\qq\label{app:eq:det-1}
&\dot{R}_t\int_{0}^{\infty} dr \, \psi_{R_t}' (\nabla^{-2}\varphi_{R_t}')
=\\
&-\dot{R}_t\int_{0}^{\infty} dr \, \psi_{R_t}'(r-R_t)
\int_r^\infty \frac{dr_2}{r_2^2} \int_0^{r_2} dr_1 \,r_1^2\, \varphi_{R_t}'(r_1-R_t)\,.\nonumber
\qqq

Integrating by parts over $r_2$, we have
\qq\label{app:eq:det-2}
(\ref{app:eq:det-1})=
&&-\dot{R}_t \int_{0}^{\infty} dr \psi_{R_t}'(r-R_t)\nonumber\\
&&\left[
\frac{1}{r}\int_0^r dr_1 \,r_1+\int_r^\infty dr_1
\right]r_1  \varphi_{R_t}'(r_1-R_t)\,.
\qqq
Concerning the first term, we make the change of variables $r\to r-R_t, r_1\to r_1-R_t$, expand in powers of $R_t^{-1}$, and go back to the original variables, to obtain
\qq\label{app:eq:det-3}
R_t\dot{R}_t &&\left\{
\psi_-\delta\phi 
+\int_{0}^{\infty} dr\, \psi_{R_t} \varphi'_{R_t}
\right\}
+\mathcal{O}(\dot{R}_t)\,.
\qqq
We now consider the second term of eq. (\ref{app:eq:det-2}). Integrating by parts over $r$ and again expanding in $R_t^{-1}$ gives
\qq\label{app:eq:det-4}
- R_t\dot{R}_t &&\left\{
\psi_+\delta\phi 
+\int_{0}^{\infty} dr\, \psi_{R_t} \varphi'_{R_t}
\right\}\,
+\mathcal{O}(\dot{R}_t)\,.
\qqq
We now add up (\ref{app:eq:det-3}) and (\ref{app:eq:det-4}) to get 
\qq\label{app:eq:det-5}
(\ref{app:eq:det-1})
= 
R_t\dot{R}_t [\delta\phi\delta\psi +\mathcal{O}(R_t^{-1})]
\qqq
which gives the desired result.

\subsection{Noise term}
We compute here the statistics of the noise entering Equation (11). To do so we start from defining 
\qq
\bar{\chi} = \int_{0}^{\infty} dr\,\psi'_{R_t} \nabla^{-2}\chi
\qqq
where we recall that $\chi$ is a Gaussian white noise with zero average and variance 
\qq
\langle \chi(r_1,t_1)\chi(r_2,t_2)\rangle  = -\frac{2D}{S_3}\nabla^2\left(\frac{\delta(r_1-r_2)}{r_1^{2}}\right)\delta(t_1-t_2)\,.\nonumber
\qqq
Because $ \nabla^{-2}$ is a linear operator, $\bar{\chi}$ is also Gaussian with zero average. We are thus left with computing its correlations. Integrating by parts, it is easy to show that
\qq
\langle \bar{\chi}(t_1)\bar{\chi}(t_2)\rangle
&=&
-\frac{2D }{S_3} \int_{0}^{\infty} dr\,
\psi'_{R_t}(r-R_t)  \nonumber\\
&&\nabla^{-2}\left(
\frac{\psi'_{R_t}(r-R_t)}{r^2}
\right)\,\delta(t_1-t_2)\,.\nonumber
\qqq
Using the solution to the Poisson equation and integrating by parts, we obtain
\qq
&&\langle \bar{\chi}(t_1)\bar{\chi}(t_2)\rangle
=
\frac{2D }{S_3 R_t}\\
&&\left\{
 2\int_{0}^{\infty} dr\,
 \psi'_{R_t}  \psi_{R_t}
 +2\psi_+\delta\psi
\right\}\delta(t_1-t_2)\,+\mathcal{O}(R_t^{-2})
\nonumber
\qqq
and hence
\qq
\langle \bar{\chi}(t_1)\bar{\chi}(t_2)\rangle
=
\frac{2D }{S_3 R_t}(\delta\psi)^2\delta(t_1-t_2)\,+\mathcal{O}(R_t^{-2})\,.
\qqq
It is then straightforward to show that the noise entering (11) is as specified in the main text.

\subsection{ Critical radius in the limit of small supersaturation }
We show here that, in the limit of small supersaturation, the critical radius in Equation (9) reduces to 
\qq\label{app:eq:Rc}
R_c = \frac{ (d-1)\sigma}{\epsilon \Delta\psi f''(\phi_1)} +\mathcal{O}(\epsilon^0,  1/(\epsilon R_t))\,.
\qqq
To do so we recall that $\phi_{\pm} = \phi_{1,2}+\alpha_{\pm}/R_t+\mathcal{O}(R_t^{-2})$ where the constants $\alpha_{\pm}$ quantify the departure of the density from the binodals due to the curvature of the interface, which can be computed as shown in~\cite{tjhung2018cluster}; their value is not needed in the following.
 Using the definitions of the transformed variables $\psi$ and $g$ we thus obtain that 
\qq
\psi(\phi_{\pm})&=&\psi(\phi_{1,2}) + \frac{\alpha_{\pm}}{R_t}\left.\frac{\p\psi}{\p\phi}\right|_{\phi=\phi_{1,2}}
+\mathcal{O}(R_t^{-2})\\
g(\phi_{\pm})&=&g(\phi_{1,2}) +\frac{\alpha_{\pm}}{R_t}\left.\frac{\p g}{\p\phi}\right|_{\phi=\phi_{1,2}}
+\mathcal{O}(R_t^{-2})
\qqq
where $|_{\phi=\phi_{1,2}}$ stands for the evaluation at $\phi=\phi_{1,2}$ and
we have used that $(\p \psi/\p\phi) (\p f /\p\phi) = \p g/\p\phi$ from the definition of $g$.

We can now expand $f'(\phi_s)\delta\psi-\delta g$ for small supersaturation and large $R_t$ to obtain 
\qq\label{app:eq:Rc_int}
f'(\phi_s)\delta\psi-\delta g = \epsilon f''(\phi_1)\Delta \psi +\mathcal{O}(\epsilon^2,\epsilon/R_t)\,.
\qqq
Plugging (\ref{app:eq:Rc_int}) into (9) we finally obtain (\ref{app:eq:Rc}).

\section{Derivation of the results in the standard variables}
While the results in this Letter were derived employing the transformed variables $\psi$ and $g$, we show here that a fully equivalent result is obtained without introducing them. To see this, we multiply (7) by $\varphi_{R_t}'$ instead of $\psi_{R_t}'$; following the same steps as in the main text, we obtain again Equations (11)-(12), 
but now 
\qq\label{eq:effective-R-equation-landscape-NV}
U(R_t) = \bar{\sigma}S_d R_t^{d-1}
\left[
1-\frac{d-1}{d}\frac{R_t}{R_c}
\right]
\qqq
where
\qq\label{eq:Rc-NV}
R_c=\frac{ (d-1)\bar{\sigma}}{f'(\phi_s)\delta\phi-\delta f}\,
\qqq
and 
\qq\label{app:eq-sigma-bar}
\bar{\sigma}=
\int dr \varphi'^2(r) \left[
K
-\frac{3}{2}\frac{2\lambda-\zeta}{d-1} \varphi_1'(r) 
-\zeta(\varphi(r)-\phi_2)
\right]\,.\nonumber
\qqq
Comparing (13) with (\ref{eq:effective-R-equation-landscape-NV}) and (9) with (\ref{eq:Rc-NV}) we conclude that $\bar{\sigma}=\sigma \Delta\phi/\Delta\psi$ and $\Delta f = \Delta g \delta\phi/\delta \psi$. Note however that  the use of the transformed variables eliminates the need to know $\varphi_1$, the correction to the profile of a flat interface, to compute the interfacial tension; $\varphi_1$ indeed does not enter in Equations (11-14).

\section{Expression for the surface tension $\sigma$}
A simple numerical procedure to obtain the binodals $\phi_{1,2}$ and the interfacial tension $\sigma$ was provided in~\cite{tjhung2018cluster}. Here, for the choice $f(\phi)=-a\phi^2/2+b\phi^4/4$, we recall how to obtain the binodals and show that $\sigma$ can be written as a single integral to be evaluated numerically. As recalled in the main text, any cubic term $c\phi^3/3$ in $f(\phi)$ can be absorbed by the shift $a\to a +c^2/(3b)$, $\phi\to \phi+c/3b$.

The binodals $\phi_{1,2}$ are obtained generalising the equality of the chemical potential and pressure, which become
\qq
&&f'(\phi_1)=\mu_1=\mu_2 = f'(\phi_{2})\label{app:eq-mu}\\
&&(\mu\psi-g)_1 = (\mu\psi-g)_2\label{app:eq-P}\,
\qqq
where the explicit expression of $\psi$ is given in the main text.
The  transformed variable $g$ is defined by $\p g/\p \psi = \p f/\p \phi$; imposing that $g= f$ when $\lambda=\zeta=0$, we obtain
\qq\label{app:eq-g}
g(\phi)&=&
-\frac{a }{\bar{\zeta}^2}+\frac{6 b }{\bar{\zeta}^4}
+a\frac{ e^{\bar{\zeta} \phi}}{\bar{\zeta}^3}\bar{\zeta} (1-\bar{\zeta} \phi )\nonumber\\
&&
b\frac{ e^{\bar{\zeta} \phi}}{\bar{\zeta}^4}
[\bar{\zeta}^3 \phi ^3-3 \bar{\zeta}^2 \phi ^2 +6 \bar{\zeta}  \phi -6  ]
\qqq
where $\bar{\zeta}=(\zeta -2 \lambda )/k$. 
Although an analytic solution for $\phi_{1,2}$ can be found only when $2\lambda=\zeta=0$ (in which case $-\phi_1=\phi_2=1$), solving Equations (\ref{app:eq-mu}) and (\ref{app:eq-P}) for $\phi_{1,2}$ is a straightforward numerical problem.

We then show that $\sigma$ can be expressed only in terms of a single integral, to be evaluated numerically. To do so, let us first observe that 
\qq\label{app:eq-sigma-single}
\sigma = 
\int_{\phi_1}^{\phi_2} d\varphi \, \sqrt{w(\varphi)}\left\{ K
-\zeta 
\left[\psi(\varphi) -\psi({ \varphi_2})\right]\right\}
\qqq
where $w(\varphi) = \varphi'^2$ is seen as a function of the density. In turn, $w(\varphi) $ can be obtained explicitly. To do so, we first observe that the profile of the flat interface $\varphi$ solves
\qq
\mu_1 = f'(\varphi) - K \varphi'' + (\lambda-\zeta/2)\varphi'^2\,.
\qqq
Multiplying by $\psi'$ and integrating in the interval $(-\infty, x)$, we obtain 
\qq
\mu_1[\psi-\psi(\phi_1)]
&=&
g-g(\phi_1) \nonumber\\
&+&\int_{-\infty}^x dx_1 \psi' [(\lambda-\zeta/2) \varphi'^2 - K \varphi'']\,.\nonumber
\qqq
Using the definition of the transformed variables to evaluate the last integral and solving for $w$:
\qq\label{app:eq-w}
&&w(\varphi)
=
\frac{2}{K} \left[
g(\varphi)-g(\phi_1) - \mu_1 (\psi(\varphi) -\psi(\phi_1)) 
\right]\frac{\p \varphi}{\p \psi}\nonumber\\
&=&
\frac{2}{K} \left[
g(\varphi)-g(\phi_1) - \mu_1 (\psi(\varphi) -\psi(\phi_1)) 
\right]e^{-\frac{\zeta-2\lambda}{K}\varphi}
\,.
\qqq
Inserting (\ref{app:eq-g}) and (\ref{app:eq-w}) into (\ref{app:eq-sigma-single}) allows us to compute $\sigma $ numerically as a single integral over $\varphi$. 


\end{document}